# ACCELERATOR OPERATORS AND SOFTWARE DEVELOPMENT

A. Miller, TJNAF, Newport News, VA 23606, USA
M. Joyce, TJNAF, Newport News, VA 23606, USA

Abstract

At Thomas Jefferson National Accelerator Facility, accelerator operators perform tasks in their areas of specialization in addition to their machine operations duties. One crucial area in which operators contribute is software development. Operators with programming skills are uniquely qualified to develop certain controls applications because of their expertise in the day-to-day operation of the accelerator. Jefferson Lab is one of the few laboratories that utilizes the skills and knowledge of operators to create software that enhances machine operations. Through the programs written by these operator-programmers, Jefferson Lab has improved machine efficiency and beam availability. Many of these applications involve automation of procedures and need graphical user interfaces. The scripting language, Tcl, and the Tk toolkit have been adopted to facilitate this type of development. In addition to automation, some operator-developed applications are used for information distribution. For this purpose, several standard web development tools such as perl, VBScript, and ASP are used. Examples of applications written by operators include injector steering, spin angle changes, system status reports, magnet cycling routines, and quantum efficiency measurements. This paper summarizes how the unique knowledge of accelerator operators has contributed to the success of the Jefferson Lab control system.

## 1 ACCELERATOR CONTROL TOOLS

Operators use the Tcl/Tk package as the basis for many of their applications because the scripting language is a great vehicle for procedural automation, centralization of tasks with GUIs, and prototyping.

### 1.1 Procedural Automation

An example of one of our automated procedures is the 100keV Steering Script. The optics in the 100keV [*]region of the Continuous Electron Beam Accelerator's (CEBA) injector makes manual steering after injector configuration changes a long, difficult process. Manually steering the electron beam in this region may take expert injector personnel more than an hour to achieve. The 100keV Steering Script was written to automatically reproduce a known good orbit using a matrix-based steering algorithm. After the response matrix is calibrated, the script can apply this matrix to make orbit corrections in less than five seconds. A new matrix is calibrated only if the optics in the region has changed, which happens infrequently. The elements of an 8X8 response matrix are found by observing the beam steering response at four locations (in the horizontal and vertical directions) with eight different magnets (four horizontal correctors, and four vertical correctors). The matrix is then inverted using a Tcl package (PBJ) specifically created by software developer Joan Sage, for this purpose. The inverse of the matrix can then be used to calculate the necessary magnet changes to achieve the desired positions at each location. The script then applies the changes. The automation of the steering in this region has saved the laboratory countless hours in injector tuning. This steering technique was later used as a prototype for the empirical locks used in the main accelerator.

The unique combination of the operator-programmer's personal experience with the long frustrating procedure and knowledge of physics and mathematics made it possible for her to conceive of and execute the automation of this common task. The end result is a tool that is now a critical part of injector setup and beam quality maintenance.

### 1.2 Centralization of Tasks

Operator-programmers also use Tcl and the Tk toolkit to create tools of convenience. These tools are not only time savers, but allow control of the system from one central interface. At Jefferson Lab, there are over 1400 trim magnets used to steer and focus the electron beam. To ensure that these correctors and quadrupoles maintain a reproducible response (current vs. BDL), the correctors are cycled through their hysteresis loops. Historically, there has been a limit on the number of trim magnets per trim rack that can be cycled reliably (the present suggested limit is three). Thus, cycling all of these magnets by hand would be very involved. Approximately 80 control screens would need to be opened, and a dedicated operator would be needed to monitor the state of the trims and cycle the next set of magnets when appropriate. A script, Hyst Area, was created by Shannon Krause (a former Jefferson Lab operator) to control the cycling from a single GUI. The script allows the user to select a combination of trim magnets to be cycled (selections can be made by area, trim rack, or ioc). The script then cycles three magnets per rack until all of the magnets have been cycled through their loops. At that time, the script audibly informs the operator that cycling is complete.

[*] This work was supported by the U.S. DOE Contact No ED-AC05-84-ER40150

The ability to control the magnet cycling from one central script ensures that magnets are not missed and frees the operator to perform other tasks instead of devoting their efforts to cycling magnets. The operator-programmer recognized the inefficiency in the manual system and designed a central interface that has made the job of many operators much easier.

### 1.3 Prototypes

Operators have created scripts as prototypes for compiled servers. These scripts run in the background at all times, collecting data and monitoring the control system. An example of a server-type application is the Beam Operations Objective Monitor (BOOM). BOOM "is an electronic time accounting system that monitors electron beam delivery from the CEBAF accelerator to specific objective points according to a set of predefined states. The information gathered is used to track beam availability to the experimental halls and the overall beam delivery history of the accelerator."[1] BOOM provides a comparison of the hours of actual beam delivered to the experimental halls versus the hours of scheduled beam delivered. It generates daily and weekly bar graphs that illustrate accelerator beam delivery performance, and the use of the beam by the experimental halls. This time-accounting information is used to measure accelerator performance per the Department of Energy performance-based contract metrics and is also a key component of the daily assessment of the accelerator's performance by lab management.

Because of his unique beam operations perspective, an operator-programmer was asked to participate in this project. This experience was very helpful, especially when it came to defining the operating parameters (i.e., when an experimental hall is "up", when the accelerator is "tuning," etc.), and then selecting the appropriate means for monitoring the necessary beam parameters.

## 2 WEB BASED INFORMATION TOOLS

In addition to applications designed primarily for the operations group to control and monitor the accelerator, some operator-developed applications are used for site-wide information distribution. For this purpose, several standard web development tools such as perl, VBScript, and ASP are used.

### 2.1 Report Forms

One of the frequently used web-based tools is the Run Coordinator Weekly Summary Reports. The run coordinator is the lead scientist for the experiment that is presently taking data in each hall. This report form was designed primarily as a way for the physicists to provide feedback on the experiment's success (or failure) during the previous week to the operations group, and to give the operations group a plan for the upcoming week. The web-based form is filled out by the run coordinator and submitted to a Microsoft Access database. The input is then manipulated by VBScript to format the data in a reader-friendly manner for presentation. Once the data has been entered into the database, it can then be retrieved for further editing and viewing using ASP web pages to retrieve the information. The benefit of having an operator design this project is realized because of the unique relationship operators have with users. Operators work twenty-four hours a day, seven days a week, fulfilling users requests and solving their beam quality issues. Knowing where the users problems are, or what they consider "good" beam quality, allows operators to more quickly identify problems before they start, or to set up the beam correctly from the beginning.

### 2.2 Electronic Logs

There are many logs that are used in the day-to-day operation of an accelerator. Historically, this logging has been via a paper logbook. Having only a single paper copy of the logbook made it difficult for on-coming operations crews to view the information while the present control room crew continued to operate the accelerator. Jefferson Lab is systematically trying to convert to electronic logging. One such conversion is the "Bugger Log." The Bugger Log is used to report mechanical or electrical systems that have been temporarily bypassed. Previously, with the paper log, information would not get entered reliably, entries were illegible, and entries would not get updated or removed. There were times that this system impaired the operations group's ability to efficiently run the accelerator; therefore, an operator converted the paper logbook into an efficient web-based tool.

From the main Bugger Log ASP web page, a bypassed system can be added or removed from the underlying Microsoft Access database by a system owner. The operations staff can then choose to view only those systems that are presently bypassed. In the future, an email message will be generated if new "buggers" are added to the database, informing the operators of the new system status. Since the operators set up the machine and are first in line to troubleshoot problems, it is important for them to know if a system has been bypassed. This helps reduce the amount of time spent investigating a system that is not the real cause of a problem.

Another log conversion that directly benefits operations is the On Call Database. The paper copy of this information was a collection of memos from the various support groups, informing the operations crew of the name of the support staff on call for that group. There are several problems with this type of information system: 1) It is hard to keep up to date, and many times the information pertained to the previous week or even month. 2) Inaccessible — it is difficult to keep this type of logbook up to date when it is located in only one place. 3) Multiple formats — each group used their own format for the on-call information. Many times, this inconsistency led to confusion as to who was the appropriate person to contact.

Because of these problems, the On-Call Database was created. Each group enters the on call information into the database through a common web form. They provide the times, dates, and names of the on-call personnel. When the operations crew has a problem and needs to contact an on-call specialist, using a web form, the operator selects the group that needs to be contacted, and a simple SQL query retrieves the name and number of the person on call at the time. The database will also warn the responsible party if the on-call list needs to be updated. The database was created to give the individual groups the flexibility to schedule on call times as they deem appropriate (i.e., eight-hour shifts for one week, twelve-hour shifts for three weeks, etc.) and the common format allows the operator to find the information they need quickly.

## 3 CONCLUSION

Operator-programmers can be a valuable resource to any software development group for a number of reasons. The operator-programmers provide extra manpower, reducing the workload on other developers. They understand the nuances of the control system and component interactions better than most, enabling them to iterate to a solution more quickly. Regular, interactive work with the accelerator allows an operator developer to test and prototype opportunistically, in line with agile development methods. The most significant asset that operators bring to their software development work is their unique role in the workplace, where they can serve as requestor, developer, and user of the software. Jefferson Laboratory has had a great deal of success with this model. The operations staff as a whole is happy with the responsiveness of their operator-programmers and the quality of the work, while the individual developers have the opportunity to expand their skills and contribute to the laboratory's success in new ways.